\documentclass[manuscript]{emulateapj}

\newcommand{\dxdy}[2]{{\frac{\partial{#1}}{\partial{#2}}}}
\newcommand{\dxdys}[2]{{\frac{\partial^{2}{#1}}{\partial{#2}^{2}}}}
\newcommand{\DxDy}

\shorttitle{Internal gravity waves in massive stars}
\shortauthors{Rogers et al.}

\begin{document}


\title{Internal Gravity Waves modulate the apparent misalignment of Exoplanets around Hot Stars}
\author{T.M. Rogers}
\affil{Department of Planetary Sciences, University of Arizona,
    Tucson, AZ, 85719}
\email{tami@lpl.arizona.edu}
\author{D.N.C. Lin}
\affil{Astronomy and Astrophysics Department, University of California, Santa Cruz, CA 95064}
\affil{Kavli Institute for Astronomy and Astrophysics and School of Physics, Peking University, China}
\email{lin@ucolick.org}
\author{H.H.B. Lau}
\affil{Argelander-Institut for Astronomie Universit Bonn Auf dem Huegel 71 53121 Bonn Germany}
\affil{Monash Centre for Astrophysics, School of Mathematical Sciences, Monash University, Australia}
\email{hblau@astro.uni-bonn.de}

\begin{abstract}
We propose that the observed misalignment between extra-solar planets and their hot host stars can be explained by angular momentum transport within the host star.  Observations have shown that this misalignment is preferentially around hot stars, which have convective cores and extended radiative envelopes.  This situation is amenable to substantial angular momentum transport by internal gravity waves (IGW) generated at the convective-radiative interface.  Here we present numerical simulations of this process and show that IGW can modulate the surface rotation of the star.  With these two-dimensional simulations we show that IGW could explain the retrograde orbits observed in systems such as HAT-P-6 and HAT-P-7, however, extension to high obliquity objects will await future three-dimensional simulations.  We note that these results also imply that individual massive stars should show temporal variations in their v sini measurements.

\end{abstract}


\section{Introduction}

Jupiter-mass planets with a few day periods (hereafter 
hot Jupiters) are found around 1-2\% of solar type stars.
At their present-day location ($<0.05$AU), {\it in situ} 
formation through core accretion \citep{pol96} is 
challenging because a1) their natal disks are relatively hot 
($ > 10^3$ K), a2) the local Keplerian velocity is $ > 10^2$ 
km s$^{-1}$ (ie much larger than the surface escape speed 
of any known planets), and a3) the tidal perturbation of their 
host stars is strong.  Consequently, b1) only a modest amount 
of refractory grains can avoid evaporation in this region, b2)
high-speed collisions among planetesimals generally lead to their 
disruption, b3) the asymptotic mass of protoplanetary embryos 
is limited, and b4) the gas accretion is quenched.

These issues have led to a paradigm shift that hot Jupiters may 
have relocated over large distances from the preferred sites 
of their assemblage \citep{il04,il10}.  The physical cause of 
planetary relocation remains a controversial issue.  A widely 
adopted ``disk-migration'' scenario attributes the cause of 
hot Jupiters' orbital evolution to their tidal interaction with
their natal disks \citep{lin96}. Their orbital decay is 
stalled when they enter into a central cavity induced by the 
host stars' magnetosphere \citep{koen91} or when their natal disk
is severely depleted.  This disk-migration scenario naturally 
accounts for the origin of multiple planets which 
are locked in mean motion resonances \citep{lp02}

Distant gas giants may also venture close to
their host stars if their eccentricity is near unity 
through dynamical excitation by the \cite{koz62} resonance \citep{wum03,ni11}, 
planet-planet scattering \citep{chat08}, or secular chaos \citep{wu11}. 
Tidal dissipation within the scattered planets would 
circularize their orbits if their periastron distance can 
reach within a few times their Roche radii \citep{rf96}. 
However, mass loss during the extremely close encounters 
also leads to either escape \citep{gui11} or 
residual cores as short-period super Earths \citep{liu12}.
This dynamical scenario predicts a pile up in the period 
distribution which does not seem to match that observed
with a control sample \citep{howard10}.

Recent observations of the Rossiter-McLaughlin (RM) effect \citep{ohta05}
indicate that the spin of many relatively massive 
and hot ($>6,300$K) main sequence stars is 
misaligned with their planets' angular momentum vector \citep{winn10,schl10}. 
This misalignment is an 
expected outcome of the dynamical scenario \citep{fab07}, although its correlation with the stellar mass 
requires assumptions on the poorly know tidal dissipation
mechanism inside main sequence stars \citep{winn11,lai12}.

The observed spin-orbit misalignment would pose a challenge 
to the disk-migration scenario for the origin of hot Jupiters
\citep{triaud10,winn11} if the host stars retain 
the angular momentum they accreted from the protostellar disks,
out of which the planets formed.  Within the observational 
uncertainty, this assumption is consistent with that observed 
in Kepler 30 \citep{so12}, but it cannot 
account for the 7 degree tilt in the Sun's spin axis with 
respect to the angular momentum vector of the planetary orbits.

Stars generally form in a turbulent cloud complex where fragments
undergo chaotic encounters as they collapse to form stars through
disk accretion.  The spin orientation of protostellar disks may 
re-orient episodically, especially during the disk formation. \cite{bate10}
suggested that this process may have led to the observed
misalignment between planets' orbital angular momentum vector 
and the spin axis of their host stars as well as differential rotation.  But this possibility may 
depend on whether 1) there is adequate mass in the disk to form 
planets, 2) there is sufficient residual gas to re-orient planets' 
orbital plane during disk depletion, and 3) there is enough coupling 
between the stars and their surrounding disks to realign the spin 
of their outer envelope. 

In this paper, we consider another possibility: that differential
spin may naturally arise in relatively massive and hot stars
even in the limit that they started with uniform spin.  
We show that internal gravity waves (IGW) generated at the interface between the convective core and radiative envelope transport angular momentum outward. These waves dissipate in a shell near the surface of the star forcing the surface (including the photosphere) to rotate differently than the bulk of the star and to vary in time.

\section{Results}
\subsection{Model Setup}
In order to study the angular momentum transport by waves in a massive star we solve the full set of hydrodynamic equations in the anelastic approximation.  These equations are solved in two-dimensional cylindrical coordinates representing an equatorial slice of the star.  A complete description of the equations and numerical method can be found in \cite{rg05}.  The reference state model is that of a 3$M_{\odot}$ star and is calculated using the Cambridge stellar evolutionary code STARS 
\citep{egg71}.\footnote{Although we use a 3$M_{\odot}$ model here we expect the results to be qualitatively representative of any star with core convection and an extended radiative envelope.}  In this model the convective core occupies the inner $\approx 14\%$ by radius and the radiative envelope extends to the stellar surface at $\approx 1.6R_{\odot}$\footnote{We do not in this first attempt include a surface convection zone for numerical efficiency.}.  Over the simulated domain, the density varies from $40 gm/cm^{3}$ in the core to $10^{-3} gm/cm^{3}$ at the surface.  Here we report mainly on the results of a model in which the convecting core is initially rotating 1.5 times faster than the radiative envelope and the average rotation is approximately ten times solar, designated MDR.  An additional model with no initial rotation, designated MUR, will be briefly discussed.  

\subsection{Angular Momentum Transport by Internal Gravity Waves}

In Figure 1 (a-e) we show the vorticity at five different times during the simulation for MDR.  We can immediately see that the small convecting region efficiently generates IGW\footnote{We use the term "wave" here loosely to mean "disturbance".} which propagate into the radiative envelope with amplitudes similar to that of convection (wave velocity amplitudes at generation are $\approx 10^{3}-10^{5}$ cm/s).  The wave generation process is ${\it extremely}$ time and location dependent.  In Figure 1a one can clearly see that waves are generated where an upwelling plume hits the convective-radiative interface.   At one time waves are generated at one main location, at another a different location and at another at multiple locations (see Figures 1b and 1c as well).  Therefore, the wave pattern seen in the radiative envelope is some combination of waves generated at various times and locations.  One can also see in Figure 1a that the wave patterns are predominantly horizontal or concentric, reflecting the fact that these waves propagate mostly horizontally because of the smallness of $\omega/N$, where $\omega$ is the wave frequency and N is the Brunt-Vaisala frequency.  The wave propagation is therefore, a spiral propagation outward.  

Once generated, the propagation and dissipation of the waves depends sensitively on the properties of the waves.  Low frequency waves are generated with higher amplitudes than high frequency waves (the generation spectrum we measure is  $A(\omega)\propto e^{\left(b_{1}\omega+ b_{2}\omega^{2}\right)}$).  However, on their propagation outward low frequency waves are dissipated by radiative diffusion more strongly than their high frequency counterparts.  The radiative diffusion acts nearly as an equalizer, so that while high and low frequency waves are generated with different amplitudes, once they reach the surface there is less dispersion.  The waves that reach the surface with the highest amplitude are those waves with high enough frequency to escape substantial radiative diffusion but low enough frequency to be generated efficiently, these tend to be waves with frequencies $\approx 10\mu Hz$ and large scales (horizontal wavenumber, k, 1-3).  It is these mid-frequency range waves that contribute the most to the angular momentum transport at the stellar surface.  Waves with this combination of frequency and wavenumber reach the surface with velocities approaching $\approx 10^{5} cm/s$, have Froude\footnote{The Froude number is a general measure of wave linearity and is defined as $w'/LN$, where w' is a velocity amplitude, L is the wavelength and N is the Brunt-Vaisala frequency.  Numbers close to unity indicate nonlinearity.} numbers close to 1 and are therefore, nonlinear.  Some evidence of wave breaking, rippled surfaces and large amplitudes, can be seen in figure 1b.  While we see some evidence of wave breaking, the actual process by which waves initially dissipate in the outer layers is not fully understood.  Several physical processes could contribute including: 1) nonlinear wave breaking 2) wave-wave interaction as waves reflect off the top boundary or 3) wave-wave interactions as waves internally reflect when propagating into a region of rapidly varying N.  

Because of the initially imposed differential rotation at the convective-radiative interface, prograde waves are doppler-shifted to higher frequencies, while retrograde waves are shifted to lower frequencies.  Therefore, radiative diffusion, being frequency dependent (the radiative damping length is proportional to $\omega^{4}$), dissipates retrograde waves ${\it slightly}$ more than prograde waves.  Consequently, upon arrival at the surface, retrograde waves have slightly lower amplitude than their prograde counter-parts and therefore, contribute slightly less to the angular momentum evolution.   This means that, although both prograde and retrograde waves reach the surface and dissipate, over time, prograde waves deposit more angular momentum and the region ${\it accelerates}$.  

Eventually the shear flow reaches an amplitude such that it represents a critical layer to many of the dominant phase speed waves \citep{bb67}.  A critical layer is defined as the position where the local angular velocity is equal to the horizontal phase speed of the wave, $c_{ph}$.  In the ray-tracing description of IGW the vertical wavenumber goes to infinity at a critical layer, therefore, the time to approach the critical layer goes to infinity.\footnote{The critical ${\it level}$ is defined as the place where the horizontal phase speed of the wave is equivalent to the mean flow speed.  As the wave is incident on the critical level there is a region below the critical level where substantial wave dissipation occurs, which is referred to as the citical ${\it layer}$.  We have not necessarily adhered to these subtle distinctions in our physical description.}  This means that any small dissipation leads to complete absorption of the wave at the critical layer.  For example, at the position labeled (1) in Figure 2b, the angular velocity is $\approx 10^{-5}$Hz, therefore, a wave with frequency 10$\mu$Hz and horizontal wavenumber one, has a horizontal phase speed $10^{-5}$Hz and experiences critical layer absorption.  Indeed, any wave in which $c_{ph}=\omega/k \geq10^{-5}$Hz will encounter the critical layer and deposit most, if not all, of its angular momentum locally.  This causes the rapid increase in the angular velocity seen between times labeled (1) and (2) in Figure 2b.  

To quantify the efficiency of angular momentum transport by IGW we turn to the evolution equation for the mean angular velocity: 
\begin{equation}
\dxdy{\overline U}{t}=-\frac{1}{\rho r}\dxdy{ r\left (\rho \overline{u_{r}u_{\phi}}\right)}{r}+\nu \left (\dxdys{\overline U}{r}+\left (\frac{2}{r}+h_{\rho}\right)\dxdy{\overline U}{r}\right)
\end{equation}
the first term on the right hand side (rhs) of (1) represents the angular momentum transport by IGW or the wave momentum flux (the second term on the rhs  of (1) represents viscous dissipation, near the surface the viscous dissipation is always at least two orders of magnitude smaller than the momentum flux by waves).  This value is $\approx 10 cm/s^{2}$ at the base of the radiative envelope and around $1 cm/s^{2}$ at the surface, although there are order 10-100 variations in time.  This can lead to ${\it rapid}$ time variation of the mean flow.  The timescale for the prograde layer to grow to its peak value is extremely rapid.  Within about 9 rotation periods, or approximately $10^{-10}\tau_{ES}$, where $\tau_{ES}$ is the Eddington-Sweet time, the angular velocity at the surface has increased by a factor of 10.  Stated another way, in the rotating frame of reference the linear velocity at the surface has increased from zero to $\approx 10^{6} cm/s$ in 9 rotation periods, consistent with the values of momentum flux given above.

As time progresses, the prograde layer migrates as continuously generated waves steepen and dissipate within the critical layer.  This causes the prograde layer to propagate ${\it toward}$ the convection zone in time (Figure 2a), similar to what occurs in the Quasi-Biennial oscillation \citep{bald01} in our own atmosphere and to the Plumb \& McEwan laboratory experiment \citep{pm78}.  Eventually the entire radiative envelope is spun up.  However, because of angular momentum conservation, the convection zone slows down (see Figure 2a). This produces a differential rotation profile that favors retrograde waves at the surface, so the process reverses and eventually, the surface will spin retrograde.  We have not been able to run this model long enough for the reversal to occur, however, one can see the slow decay of the prograde shear in Figure 2b as well as retrograde pockets in Figure 1e.  At the current rate of deceleration (which is substantially slower than the rate of acceleration since no retrograde critical layer has yet formed) the prograde radiative envelope would take approximately ${\it a}$ ${\it few}$ ${\it hundred}$ ${\it rotation}$ ${\it periods}$ ${\it to}$ ${\it reverse}$, or $10^{-7}\tau_{ES}$, which astronomically speaking, is still an extremely short timescale.  The growth and reversal timescales are inversely proportional to the wave momentum flux \citep{plumb77}.  The reversal timescale is longer than the initial growth time because the convection zone contains most of the mass, therefore, when it slows down the differential rotation between it and the surrounding radiation zone is weaker than the initially imposed differential rotation.  Hence, the amplitude variation between prograde and retrograde waves at the surface is reduced and the net momentum flux is lower.  

In Figure 1(f-j) we show the vorticity for the model MUR, in which there is no initial rotation.  Here we clearly see dynamics similar to that seen in Figure (a-e) but the initial surface shear that develops in this model is ${\it retrograde}$.  In a model with uniform initial rotation the sense of the initial surface shear layer is random as there is no preferred direction.  Otherwise, this model evolves similarly to the one previously described with the development of a critical layer, rapid momentum transport and momentum fluxes similar to those quoted above for MDR.  Although we have only presented these two models we note that this is a very robust mechanism as a surface shear flow has developed in ${\it all}$ models we have run, whether initially uniformly or differentially rotating.  The amplitude of the shear and its growth rate depend on the model details but the time dependent shear development is rigorous.  The two models presented demonstrate the large variation in amplitude and sign that surface flows on hot stars can experience.


\section{Discussion}
In the above models we have shown that IGW can efficiently transport angular momentum in stars with convective cores and extended radiative envelopes.   This transport causes the angular velocity profile to vary substantially in radius and time.  In the model presented the angular velocity at the surface changed by a factor of 10 in a timescale minuscule compared to other relevant timescales.  In ${\it all}$ cases run the angular velocity changes in time.  This has significant consequences for observations of massive star rotation, as these models indicate that the rotation of the surface layers is not necessarily representative of the rotation of the entire star.  They also indicate that the surface rotation of the star is time dependent.  
This is a robust process, which happens in our own atmosphere \citep{bald01}, has been demonstrated experimentally \citep{pm78}, and shown here numerically.  It likely occurs in any star with a convective core and extended radiative envelope and has many interesting consequences.  

This mechanism provides a natural explanation for why mis-aligned planetary systems tend to be around hot stars.  In hot stars which have convective cores and radiative envelopes IGW transport sufficient angular momentum to cause the surface of the star to rotate differently than it did when the planetary system was formed.  Although our 2D models only strictly explain retrograde systems we view this work as a proof of concept and application to other high-obliquity objects will await future 3D simulations.\footnote{We should note that three-dimensional simulations of this sort have not been conducted and are highly computationally demanding as they would require substantial vertical resolution in order to resolve critical layers.}  These simulations imply, at the very least, that the equatorial angular velocity at the surface could be time-dependent as the IGW-driven shear flow varies in time.  This physical effect would lead to variations in measured RM effect.  Multiple observations of the exoplanet system XO-3 by \cite{hebrard08}, \cite{winn09} and \cite{hirano11} reveal an unexplained variation in spin-misalignment, which could be interpreted as a 30\% variation in equatorial velocity over a year.  Multiple observations of the RM effect in the same hot-star system over time could provide an important test.  Similarly, inclination measurements of massive stars over time should show variations.  
 
One way to distinguish this process, in which the mis-alignment is rooted in angular momentum variations in the star alone, from one in which the planets themselves are mis-aligned, would be the observation of multiple planets in the same system with similar mis-alignments.  This would be difficult to produce in a scattering-circularization scenario.  Recent observations of the multi-planet system Kepler-30 \citep{so12} around a solar type star does not provide a test of our model.  Observations of multi-planet systems around a hot star would provide a crucial test.

This wave-mean flow interaction could have several other interesting implications.  The surface angular velocity achieved in this model is unstable to the shear instability, likely resulting in turbulence and substantial mixing of species.  The angular velocities achieved here are about 2\%$\Omega_{c}$ and higher angular velocities are likely possible.  These amplitudes, combined with the short timescale of angular velocity variation could explain the observed BE class of stars.  

How this mechanism operates in three-dimensions (3D) is not clear.  In our own atmosphere IGW driven shear flows are limited to equatorial regions because the IGW forcing is offset by the Coriolis force at higher latitudes \citep{bald01}.  Although the analysis may be slightly altered in massive stars (as hydrostatic and geostrophic balance do not apply), it is highly likely these stars have latitudinal differential rotation, which should be accounted for when modeling the RM effect.  More detailed analysis of latitudinal and temporal variation in surface angular velocity await more sophisticated simulations.  

We suggest several observational tests of this scenario.  RM observations should be analyzed assuming some latitudinal differential rotation.  Repeated observations of the RM effect associated with the same hot star system would be a critical test and finally, observation of a multiple-planet system around a hot star in which all of the planets had the same mis-alignment would provide strong support for this model.
\bibliographystyle{apj}
%

\acknowledgments
We are grateful to G. Glatzmaier, K.B. MacGregor and Dan Fabrycky for helpful discussions.  Support for this research was provided by a NASA grant NNG06GD44G.  T. Rogers is supported by an NSF ATM Faculty Position in Solar physics under award number 0457631.   D.N.C. Lin was supported by NASA (NNX08AM84G), NSF (AST-0908807) and University of California Lab Fee. Computing resources were provided by NAS at NASA Ames. 
 
\email{aastex-help@aas.org}.

\begin{figure}
\epsscale{1.00}
\plotone{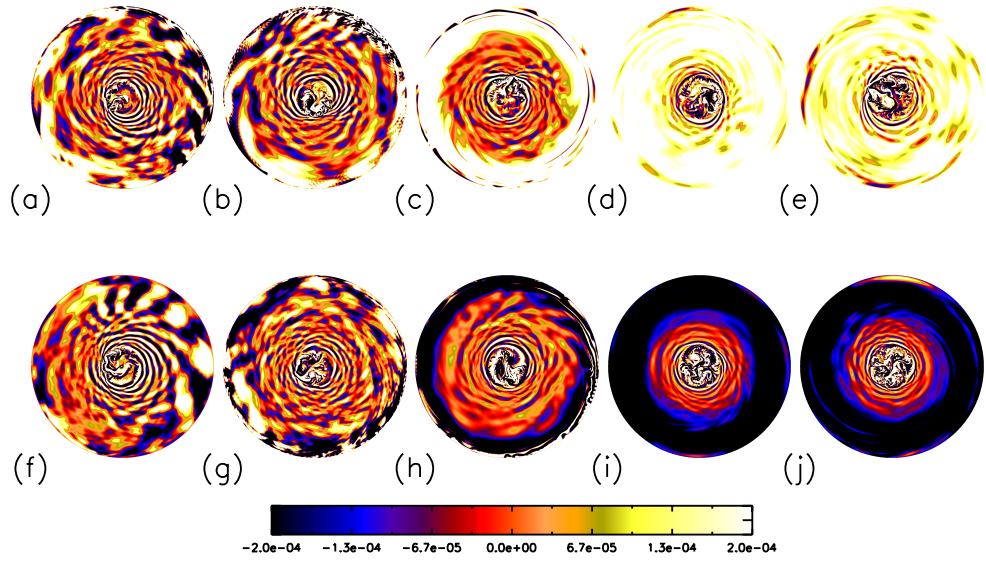}
\caption{(a-e) Time snapshots of vorticity at five different times during the simulation MDR, corresponding to (a) $4.8\times 10^{5}$s, (b) $1.76\times 10^{6}$s, c) $4.6\times 10^{6}$s, (d) $2.2\times 10^{7}$s and (e) $4.4\times 10^{7}$s. (f-j) Time snapshots of vorticity at five different times during the simulation MUR, corresponding to (f) $2.57\times 10^{6}s$, (g) $2.66 \times 10^{6}s$, (h) $5.08\times 10^{6}s$, (i) $1.15\times 10^{7}s$ and (j)$1.23 \times 10^{7}s$.  White represents positive vorticity, while black represents negative vorticity.}
\label{fig:vortsnap}
\end{figure}
\begin{figure}
\epsscale{1.00}
\plotone{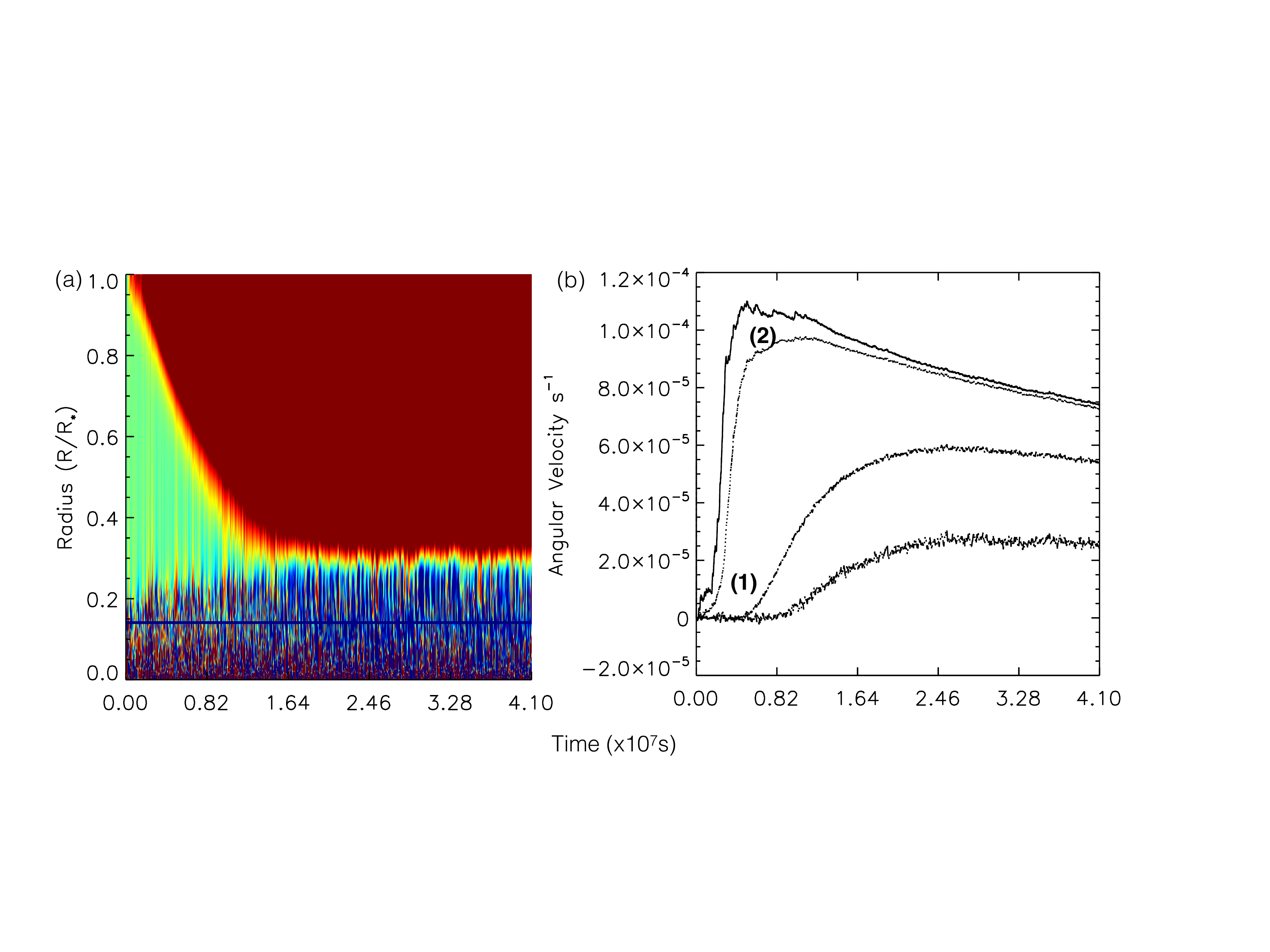} 
\caption{(a) Angular velocity as a function of time (x) and radius (y) for model MDR.  Red represents prograde flow while blue represents retrograde flow.  The blue horizontal line represents the top of the convection zone.  We see a prograde layer form initially at the stellar surface and then propagate toward the convection zone in time.  Initially the convection zone and overshoot layer show both prograde and retrograde flow.  However, as the radiative envelope is spun up, by angular momentum conservation the convection zone must slow, as is seen. (b) Line plot of angular velocity in time for radii (moving from top to bottom) 0.98, 0.86, 0.45 and 0.17 of the stellar radius.  Angular momentum deposition occurs by critical level absorption between time labeled (1) and (2).  The angular velocity has increased by a factor of 10 in just 9 rotation periods.  Subsequent decay represents the gradual reversal process as retrograde waves now reach the surface with slightly higher amplitude.}
\label{fig:vortsnap}
\end{figure}

\end{document}